\documentclass[pre,twocolumn,showpacs,preprintnumbers,amsmath,amssymb,floatfix]{revtex4}

\usepackage{graphicx}
\usepackage{dcolumn}
\usepackage{bm}
\usepackage{enumerate}

\usepackage[normalem]{ulem}  
\usepackage[dvips]{color}

\begin{document}


\title{Experimental analysis of lateral impact on planar brittle material}

\author{F.P.M. dos Santos$^{1,2}$}
\email{filipepms@gmail.com}
\author{V.C. Barbosa$^{1}$}
\email{valmar@if.ufrj.br}
\author{R. Donangelo$^{1,3}$}
\email{rdonangelo@gmail.com}
\author{S.R. Souza$^{1,4}$}
\email{srsouza@if.ufrj.br}
\affiliation{$^1$Instituto de F\'\i sica, Universidade Federal do Rio de Janeiro
Cidade Universit\'aria, \\CP 68528, 21941-972, Rio de Janeiro, Brazil}
\affiliation{$^2$ CEFET Qu\'\i mica de Nil\'opolis - Av. L\'ucio Tavares 237, 26530-060, Nil\'opolis, Brazil}
\affiliation{$^3$Instituto de F\'\i sica, Facultad de Ingenier\'\i a, \\
CP 30, 1100 Montevideo, Uruguay}
\affiliation{$^4$Instituto de F\'\i sica, Universidade Federal do Rio Grande do Sul,\\
Av. Bento Gon\c calves 9500, CP 15051, 91501-970, Porto Alegre, Brazil}

\date{\today}

\begin{abstract}
The fragmentation of alumina and glass plates due to lateral impact is studied.
A few hundred plates have been fragmented at different impact velocities and the produced fragments are 
analyzed.
The method employed in this work allows one to investigate some geometrical properties of the
fragments, besides the traditional size distribution usually studied in former experiments.
We found that, although both materials exhibit qualitative similar fragment size distribution function, 
their geometrical properties appear to be quite different.
A schematic model for two-dimensional fragmentation is also presented and its predictions are compared to
our experimental results.
The comparison suggests that the analysis of the fragments' geometrical properties constitutes a more stringent
test of the theoretical models' assumptions than the size distribution.
\end{abstract}

\maketitle

\section{\label{sec:introuduction}Introduction\protect}
\label{sect:introduction}
The fragmentation process is a common phenomenon that is found both in natural
\cite{galaxies,asteroids,fragGeoSystems,Korsnes,iceRussians,BettyPhysRep2005,molecules}
and industrial processes \cite{graniteExplosion,fragComminution,acousticSignals,dynFrag2D}, 
and takes place on scales ranging from the collision of 
galaxies \cite{galaxies} and asteroids \cite{asteroids} to the breakup of heavy nuclei \cite{BettyPhysRep2005}.
The underlying physics of the fragmentation phenomenon is very different for macroscopic and microscopic systems,
since quantum effects have to be considered in the latter, and also because the mechanisms that lead to their breakup
are not the same.
In this work, we focus on the macroscopic systems.
From both the economic and academic points of view, it is important to understand the
mechanisms that govern this process and, based on them, build models that
allow predictions for different scenarios.

As discussed in Ref.\ \cite{abramowitz}, brittle materials such as glasses and ceramics 
exhibit no macroscopic plastic deformation as they are subject to tensile loading 
capable of producing a stress level greater than their limiting elastic limit.
Thus, once this stress threshold is reached, the fragmentation process immediately starts. 
The first systematic studies of brittle fracture using statistical arguments were 
performed about 60 years ago by Weibull \cite{weibull},
who introduced a probability distribution function (named Weibull distribution) to describe the fragment size
distributions.

Recently, this subject has regained a great deal of attention since Oddershede {\it et al.} \cite{ODDERSH} showed that
the size ($s$) distribution of the fragments produced in the breakup of brittle objects is given by a power law 
$s^{-\beta}$ which is fairly independent of the specific material employed.
It is essentially determined by the morphology of the fragmenting object and the power law exponent $\beta=1.63$ was found
for spherical gypsum balls, $\beta=1.08$ for thin gypsum disks, and $\beta=1.05$ for gypsum rods.
These authors showed that this scaling law is observed for fragments whose dimensions are smaller than the smallest dimension of the
original object \cite{ODDERSH}.
Subsequent investigations made in Ref.\ \cite{GLASSrod}, where glass rods were dropped from increasing heights, suggest that the
power law exponent rises from $\beta\approx 1.2$ to 1.5 as the violence of the impact increases (larger heights).
A similar experiment has been reported in Ref.\ \cite{fragJapanese}, where 
a sandwich of thin glass or plaster plates, inserted between stainless steel plates, was hit by
an iron projectile, which was dropped on the target at normal angles to its surface.
The fragment distribution is also given by a power law whose exponent increases from $\beta=1.5$ to $1.7$ according to the
violence of the impact.
Although these studies agree qualitatively, the observed power law exponents are somewhat different.
Qualitatively different results were obtained in Ref.\ \cite{PLATESprl} where the mass distribution of fragments produced in the
fragmentation of thick plates of dry clay were described by two power laws of exponents and $\beta_S=1.5 - 1.7$ for small fragments
and $\beta_L=1.1 - 1.2$ for the larger ones.
These authors suggest that the exponents are associated with the dimensionality of the fragmenting object and the length of
the fragments.
They also investigated the problem using models based on very different pictures for the fragmentation process and found that both lead
to the same mass distributions.
This finding strongly suggests that the study of other observables, besides the mass/size distribution, is necessary to unveil the
underlying physics of the fragmentation process.
This conclusion is one of the motivations for the present work.

Other recent experimental studies investigated the properties of fragmentation in different scenarios.
For instance, the breakup of closed thin shells due to impact and also as a result of the explosion of combustible mixture
has been studied in Ref.\ \cite{HERRegg}.
For practical reasons, eggs have been used in these experiments.
The authors found that the mass distribution follows a power law, in both cases, whose exponent $\beta=1.35\pm 0.02$ lies
between $\beta_L$ and $\beta_S$ observed in Ref.\ \cite{PLATESprl} and mentioned above.
On the other hand, studies on the fragmentation of mercury drops \cite{drops,dropsPRL}, due to the fall from a fixed height on
a hard surface, show that the fragment size distribution is fairly well described by a power law of exponent $\beta\approx 1.1$
over a wide range of sizes \cite{drops}.

The role played by the constraints imposed to the fragmenting system has been investigated in Ref.\ \cite{SPAGHETTI}, where
slender brittle rods, made up of dry spaghetti (besides other brittle materials), have been kept fixed at one end and axially
impacted at the other.
The corresponding mass distribution exhibits bumps around $\lambda/2$ and $\lambda/4$, where $\lambda$ is the
preferred wavelength for the buckling instability.
This issue was, to some extent, also investigated in Ref.\ \cite{fragJapanese} where the paster plates were laterally bombarded by
a high velocity projectile and $\beta$ was found to lie within 1.1 and 1.3, which contrasts with the values mentioned above
when the plates are hit on their surfaces.
There have also been experimental studies that focused on the internal details of the fractures, such as those reported in
Ref.\ \cite{fragAlex2007}, where the statistical distribution function for the height fluctuations along the fracture length was
carefully examined and found to be Gaussian.

Owing to the great complexity of the fracture process \cite{dynCrack2003,reviewFragmentation,dynCrackYoffe,dynCrack1},
several schematic models have been proposed to describe the phenomenon (see
\cite{dynCrackYoffe,dynCrack1,HerrmannGranular1995,HerrmannGranular2008,reviewFragmentation,HERRdisk1,HERRdisk2,mestrado,Teo1prl,Teo2prl,fragModelChile2006,drops,chineseFragModel,moldyn,SIM_2D}
and references therein).
Some of them try to incorporate microscopic information \cite{dynCrackYoffe,dynCrack1,HerrmannGranular1995,HerrmannGranular2008,SIM_2D}
whereas the majority of the approaches are minimalistic models which uses as few parameters as possible to describe the fragmentation
process \cite{HERRdisk1,HERRdisk2,mestrado,Teo1prl,Teo2prl,fragModelChile2006,drops,chineseFragModel,moldyn}.
Due to the very small time scale of the fracture dynamics \cite{dynCrack2003}, most of the experimental information is usually
restricted to the final state properties of the system.
Therefore, it is very difficult to single out the appropriate scenario from these models since most of them make predictions which
agree reasonably well with the available experimental data.

In this work we present the results of an experiment from which we extract further observables associated with the fragments' geometry
and size, besides their mass distribution.
By providing more detailed information on the fragments produced in the breakup of brittle material, we intend to make it possible to
distinguish between the different scenarios assumed in the theoretical models or, at least, to constrain the range of their free
parameters.
We study the fragmentation due to lateral impacts on thin plates of alumina and glass, so that our experiment is similar to one of the
measurements made in Ref.\ \cite{fragJapanese}, mentioned above.
As described below, we have been particularly careful in controlling the impact velocities, in order to minimize effects associated
with mixing the energy deposited into the system.
Great care has also been taken in order to ensure that the plates are always hit along their lateral side, in order to minimize
any bias due to angle mixing.

The remainder of this paper is organized as follows.
Our experimental setup is described in detail in Sect.\ \ref{sect:expSetup} and the model devised to interpret the experimental
results is presented in Sect.\ \ref{sec:model}.
The results are discussed inc Sect.\ \ref{sec:results} and the main conclusions are summarized
in Sect.\ \ref{sec:conclusions}.

\section{\label{sect:expSetup}The experimental setup \protect} 
We have used 232 square plates (10cm $\times$ 10cm) to study the properties of the breakup of thin brittle plates 
due to lateral impacts.
Two different materials have been considered: alumina and glass.
The samples have different microscopic structure, but both have noncrystalline atomic arrangement. 
The alumina targets were manufactured by Coorstek Inc. and the glass plates by GoesVidros. 
None of the plates has been previously synterized.
Their physical properties, as well as the number of analyzed plates, are given in Table \ref{tab:brittleProp}.

\begin{figure}[t]
\includegraphics[width=8.5cm,angle=0]{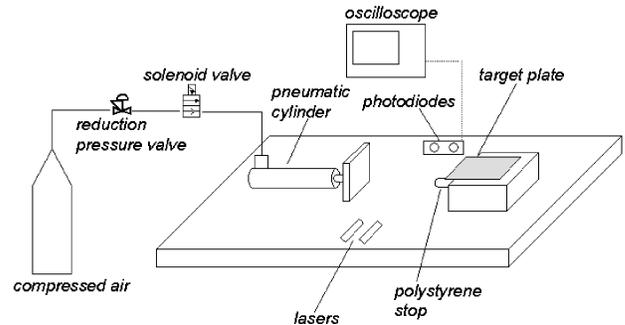}
\caption{\label{fig:apparatus} Schematic illustration of the apparatus used in this work to study the fragmentation of brittle plates.
For details, see the text.}
\end{figure}

\begin{table}[ht]
\caption{\label{tab:brittleProp}Properties of the brittle plates used in this work.}
\begin{ruledtabular}
\begin{tabular}{ccccc}
Material & Thickness & Width & Density & Quantity\\
         & (mm) & (mm) & (g/cm$^3$) & \\
\hline
Alumina & 0.5 & 100.0 & 0.380 & 156\\
Glass & 1.0 & 100.0 & 0.246 & 76
\end{tabular}
\end{ruledtabular}
\end{table}

A plate is laid down on a flat surface and it is hit laterally by a piston.
No constraints are imposed on the plate, so that it can be freely scattered.
A schematic illustration of the apparatus is shown in Fig.\ \ref{fig:apparatus}.
It consists of a compressed air cylinder which is connected to a reduction pressure valve.
The latter allows one to suitably fix the pressure in the pneumatic cylinder connected at its end.
In order to ensure uniform impacts along the plate's length,
a 12cm$\times$5cm$\times$2cm (length$\times$height$\times$width) steel block
is attached to the external end of the piston rod.
Since the length of this block (12cm) is larger than that of the plates (10cm) the impact is very homogeneous.
The basis for the targets is fixed on an iron plate.
To it is attached a polystyrene shock absorber bar, which is intended to stop the pneumatic 
cylinder before it reaches the piston stroke, thus preserving the physical integrity of the apparatus. 
The compressed air is injected into the pneumatic cylinder through the activation of 
a high-speed electrical valve, allowing the steel block to accelerate and reach the 
target plate, before being stopped by the polystyrene stop.

\begin{figure}[bt]
\includegraphics[height=6.0cm,angle=0]{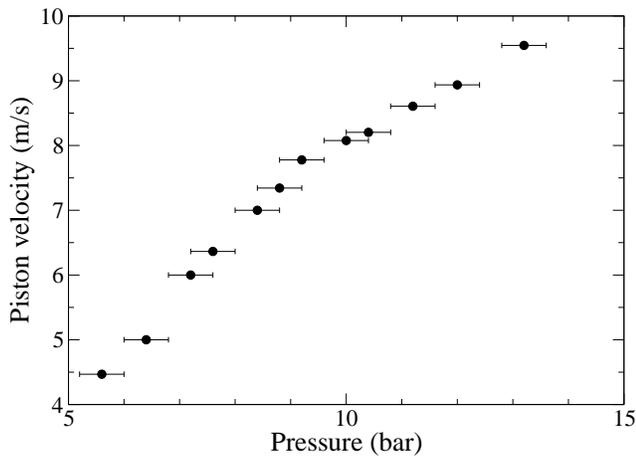}
\caption{\label{fig:calibration} Relation between the applied pressure
and the velocity of the collision block.}
\end{figure}

The velocity of the block impinging on the target is measured using two photodiodes 
located 1.2cm apart from each other.
Each of them is illuminated by a laser.
As the light on either of the photodiodes is interrupted by the passage of the block,
signals are transmitted to a digital oscilloscope, thus allowing the measurement of the block 
speed as a function of the applied pressure.
The calibration curve obtained with the apparatus is presented in Fig. \ref{fig:calibration}.

The minimal velocity necessary to break the samples depends on the material used as target.
For our experimental setup, the threshold for the alumina plates is approximately 8.1 m/s 
and it is 6.4 m/s for the glass ones.
Below these values, some of the plates occasionally break, but produce very few fragments.
If a plate does not break, it is not used again as it may have suffered internal damages.

Since the targets are not fixed, they move freely after having been hit by the piston.
In order to confine the produced fragments and at the same time preventing secondary fragmentation due to further impacts with the surroundings,
we have placed the system inside a soft plastic enclosure with a silk bag at its right end.  
This part of the apparatus is not depicted in Fig.\ \ref{fig:apparatus}.

The produced fragments are placed on a high resolution scanner.
This provides detailed images from which one determines their geometric properties.
We have been particularly careful in placing the fragments on the scanner so as to prevent
them from touching each other.
For the alumina plates, we found that black and white images of 600 dpi resolution are suitable for our purposes as they allow one to
identify fragments whose sizes are on the order of $A_{\rm cut}=0.18$~mm$^2$, i.e. $A_{\rm cut}=1.80\times 10^{-5}$ smaller than the original
objects ($A_0$).
The glass plates were scanned as 256 color bitmaps at 200 dpi resolution.
Therefore, the smallest fragment's area studied in this case corresponds to $A_{\rm cut}=1.61$~mm$^2$.
The fragments are identified by simply counting the contiguous active pixels on the scanned image,
similarly to what is done in standard cluster recognition algorithms used in percolation theory.
We stress that the analysis performed in this work provides much more information than the
traditional ones which focus on the fragments' masses.

\section{\label{sec:model}A schematic model for plaque fragmentation\protect} 
Before presenting the fragmentation data obtained with the apparatus described above, we introduce a schematic
fragmentation model which will be of help in the interpretation of our experimental results.

\begin{figure}[bt]
\includegraphics[width=8.5cm,angle=0]{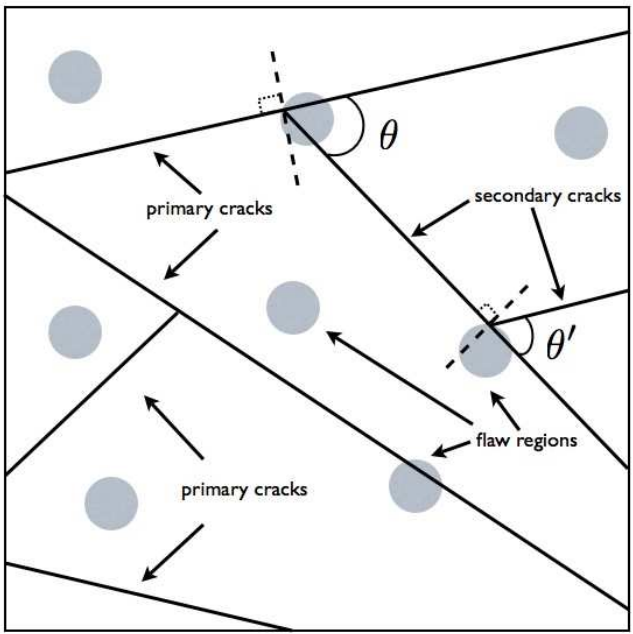}
\caption{\label{fig:model} Illustration of the propagation of the cracks in the framework of the schematic fragmentation
model introduced in this work.
The initial cracks start at the left side of the plate.
The definition of the bifurcation angle $\theta$ is also depicted in this figure.
For details, see the text.}
\end{figure}

As mentioned in the previous section, many models have been developed to describe fragmentation phenomena in different scenarios 
\cite{dynCrack1,dynCrackYoffe,HerrmannGranular1995,HerrmannGranular2008,reviewFragmentation,HERRdisk1,HERRdisk2,mestrado,Teo1prl,Teo2prl,
      fragModelChile2006,drops,chineseFragModel,moldyn}.
The confrontation of these models with the experimental results presented below is important in order to clarify the
essential physical ingredients involved in the fragmentation process.
However, due to the large number of models available in the literature, this is beyond the scope of the present work.
We therefore devised a schematic two-dimensional Monte Carlo model, depicted in Fig.\ \ref{fig:model}, inspired in the beautiful
experiment reported by Xu and Rosakis \cite{dynCrack2003}, which we apply to interpret our experimental results. 
Although we keep it as simple as possible, we introduce the main ingredients we consider relevant to the process:
\begin{enumerate}[(i)]
  \item At the beginning of the process, cracks are created at one of the lateral sides of the square plates of area $A_0=1$ and
  their propagation directions are randomly selected.
  This number $N_c$ of initial cracks is closely related to the violence of the impact and is a parameter of our model;
  \item The propagation of the cracks start simultaneously and any of them stops only if its course is interrupted
  by another crack or if one of the borders is reached.
  \item A number $N_f$ of flaw regions are randomly placed over the surface of the plate.
  The number $N_f$ is sampled from a Poisson distribution with an average value $\langle N_f\rangle$, which is a parameter
  of the model. The flaw regions are considered to be circles, all of radius $R$, which is another parameter of our model;
  \item When a crack enters into a flaw region, a new branch is created with probability $P_c$.
  Its propagation direction is uniformly sampled between $-\pi/2$ and $+\pi/2$, with respect to that of the initial crack.
  The latter, continues its course;
  \item
  In order to minimize the number of free parameters of the model, we argue intuitively that
  the number of initial cracks $N_c$ is closely related to the violence of the impact, as well as $P_c$.
  Although it is difficult to determine the relationship between $N_c$ and $P_c$, any reasonable function which
  gives $N_c=0$ for $P_c\rightarrow 0$ and leads to large values of $N_c$ for $P_c\rightarrow 1$,
  should provide equivalent qualitative results.
  For simplicity, we adopt $N_c=-10\ln(1-P_c)$.
\end{enumerate}

\noindent
The average number of flaws $\langle N_f\rangle$ is a parameter of the model which is associated with the brittleness of the material.
Therefore, different values of $\langle N_f\rangle$ and $R$ could be used for distinct materials.
For simplicity, we fixed $\langle N_f\rangle = 10000$ and $R=0.0005$ for both alumina and glass plates.
Then, the probability of creating a new crack $P_c$ is the only parameter that we vary according
to the violence of the impact and the material of the plates.
Owing to the stochastic nature of the model, ten thousand events are run for each case studied below.

\begin{figure}[t]
\includegraphics[width=8.0cm,angle=0]{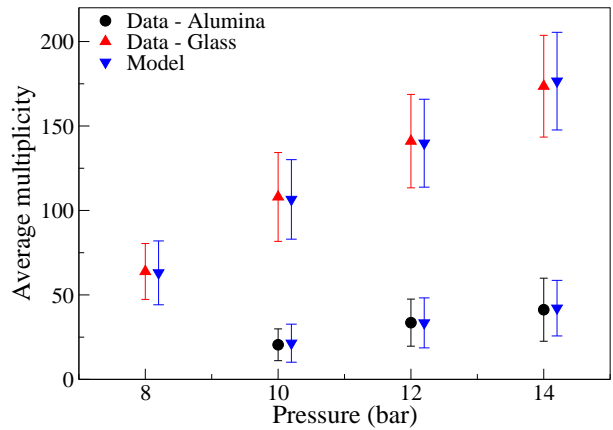}
\caption{\label{fig:mult} (Color online) Fragment multiplicity as a function of the initial cylinder pressure.}
\end{figure}

\section{\label{sec:results}Results\protect}
The average fragment multiplicity is displayed in Fig.\ \ref{fig:mult} as a
function of the pressure for both alumina and glass plates.
For each event, fragments of area smaller than $A_{\rm cut}$ are not considered as discussed in Sect.\ \ref{sect:expSetup}.
For consistency, the same procedure is employed in the theoretical calculations.
The model results, depicted by the upside-down triangles, are slightly shifted to the right in order
not to overlap with the data.
All the error bars shown in this figure correspond to the standard deviation, i.e. they are associated with the width of the distribution.
The model parameter $P_c$ was adjusted for each pressure used in the experiment.
In the case of the alumina plates, we used $P_c=0.329,\, 0.393$ and $0.425$ for $P=10,\, 12$ and $14$~bar,
respectively.
For the glass plates, we adopted $P_c=0.53,\, 0.67,\, 0.76$ and $0.85$ for $P=8,\, 10,\, 12$ and $14$~bar, also respectively.
These results show that the average fragment multiplicity increases steadily as a function of the pressure.
One also sees that it is much higher for the glass plates than in the case
of the alumina targets and that, as anticipated in Sect.\ \ref{sect:expSetup}, the fragmentation
threshold is much lower in the former case.
The large variance values, represented by the error bars, reveal that, for a given pressure, the fluctuations are fairly large.
One should note that, although $P_c$ has been adjusted to reproduce the average multiplicity, the model also correctly predicts the
large variances observed experimentally.
This is an intrinsic property of the model. 

\begin{figure}[bt]
\includegraphics[width=7.0cm,angle=0]{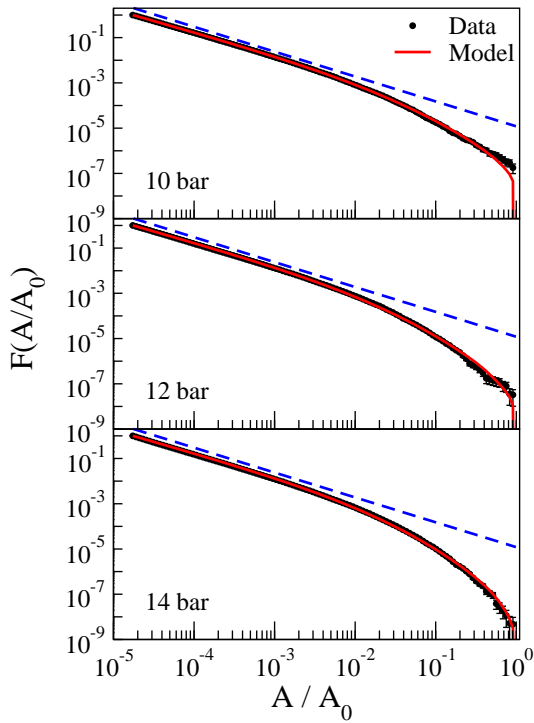}
\caption{\label{fig:areaDistA} (Color online) Area distribution for 
alumina plates at different pressures.
The dashed line is a power law of exponent $\beta=1.1$.
For details, see the text.
}
\end{figure}

We now turn to the area distribution and show in Fig.\ \ref{fig:areaDistA}, for the alumina plates, the values of $F(A)$, defined as
\cite{ODDERSH}

\begin{equation}
F(A)=\frac{1}{A}\int_{A}^{\infty}n(A^{\prime})dA^{\prime} \;,
\label{F(m)}
\end{equation}

\noindent
where $n(A)dA$ is the number of fragments with area between $A$ and $A+dA$.

The dashed line shown in this figure represents the power law fit $F(A)\propto A^{-\beta}$.
The results show that the experimental data can be fairly well approximated by this function
over about three decades with $\beta=1.1$.
This exponent is in agreement with some of the results reported previously \cite{fragJapanese,drops}.
One observes a steeper drop of $F(A)$ at large areas, which becomes
more important as the violence of the impact increases.
Nevertheless, the fragment distribution associated with not too large fragments
($A / A_0\lesssim 10^{-2}$) remains essentially unchanged within the pressure range studied in this
work.

The agreement of the model results, depicted in this figure by the full lines, with the
experimental data is very good.
Small discrepancies are observed only for large areas, but they are compatible with
the experimental uncertainties in this area region.

\begin{figure}[bt]
\includegraphics[width=7.0cm,angle=0]{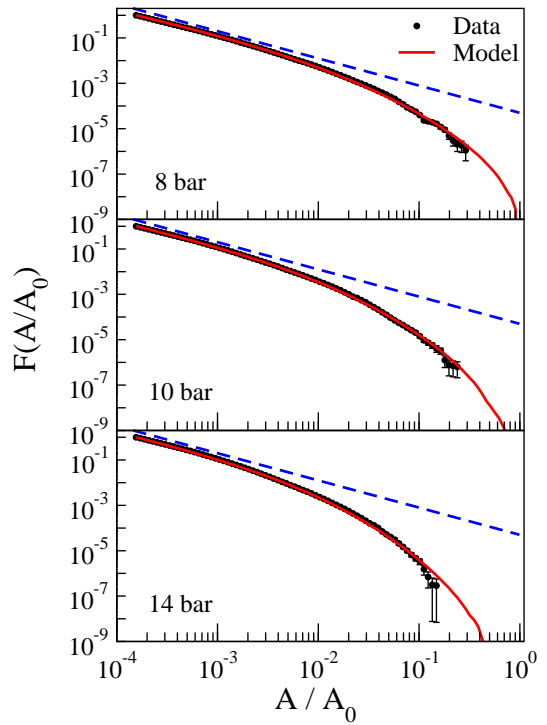}
\caption{\label{fig:areaDistG} (Color online) Same as Fig.\ \ref{fig:areaDistA} for glass plates.
The dashed line corresponds to a power law whose exponent is $\beta=1.2$.
For details, see the text.
}
\end{figure}

Qualitatively similar results are observed for the glass plates, as is shown in Fig.\ \ref{fig:areaDistG}.
The exponent of the power law function used in this case is slightly larger than that obtained above
and it corresponds to $\beta=1.2$.
The range over which this function remains a good approximation to the actual behavior observed
experimentally is considerably smaller than in the case of the alumina plates.
The suppression of large areas is much more pronounced in the present case.
This is in qualitative agreement with the results shown in Fig.\ \ref{fig:mult}, since the fragment multiplicity is
much larger in the case of the glass plates than for the alumina ones.

The model also reproduces the area distribution very well in this case.
However, as mentioned in Sect.\ \ref{sect:introduction}, it was demonstrated in Ref.\ \cite{PLATESprl} that
different assumptions for the fragmentation mechanisms may lead to equivalent mass distributions.
Therefore, the description of $F(A)$ should be regarded as a preliminary selection criterion and
further comparisons with the experimental data should be made in order to validate any model.
In the following, we examine some size correlations in order to seek for further information on the
fragmentation process.

\begin{figure}[bt]
\includegraphics[width=7.0cm,angle=0]{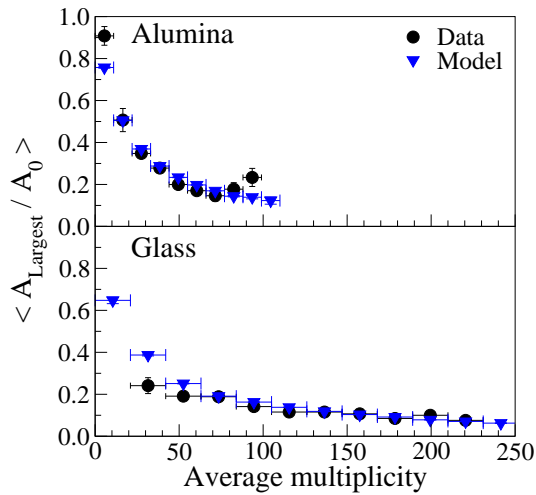}
\caption{\label{fig:am} (Color online) Average value of the largest area within each event as a function of
the multiplicity.
In the case of the alumina plates,
multiplicity bins of 11 units were used whereas 21 units were employed for the glass plates.
}
\end{figure}

The average value of the largest area, $\langle A_{\rm Largest}\rangle$, within each event is displayed in Fig.\ \ref{fig:am} as
a function of the fragment multiplicity for the alumina and glass plates.
In both cases, $\langle A_{\rm Largest}\rangle$ decreases as a function of the fragment multiplicity for not too high
multiplicities.
In the case of the alumina plates, one observes a slight increase at the highest multiplicity values.
This clearly contrasts with the behavior observed in the glass plates.
We believe that this deviation is due to the poor statistics of the very high multiplicity events since no anomalies were
observed in these events.
Nevertheless, it could also be explained by a geometrical preference for the creation of the fragments.
Since this information is not available in the present measurements, this hypothesis cannot be investigated at this
moment.
The predictions of the model agree fairly well with the alumina data for multiplicities smaller than 70 - 80.
It decreases smoothly as a function of the fragment multiplicity.
In the case of the glass plates, small discrepancies are observed at the lowest multiplicities, where the model tends to
overpredict the size of the largest fragment  whereas it underpredicts this observable for the
alumina plates.
This is an indication that, in spite of the very good overall agreement observed above, the details associated with the
fragment distribution might not be correctly reproduced by the model.

Important information on the fragmentation pattern may be obtained through the Dalitz plot \cite{Dalitz}, which is based on the
properties of the largest fragments.
More specifically, this is constructed by calculating

\begin{equation}
\chi_i=A_i\Big/\sum_{k=1}^3A_k\;,\;\;k=1,2,3,
\label{eq:Dalitz}
\end{equation}

\noindent
where $\left\{A_k\right\}$ corresponds to the area of the largest three fragments within each event.
The quantity $\chi_i$ represents the perpendicular distance to the k-{\it th} side of an equilateral triangle of height 1,
into which a point associated with a given event is plotted.
By construction, all the points lie inside the triangle.
Due to geometrical constraints, an event point is univocally defined by the pair of distances $(\chi_i,\chi_j)$,
$i\ne j$.
The indices $\left\{k\right\}$ are randomized in each event in order to eliminate artificial structures,
i.e., the labels assigned to each of the three largest fragments are shuffled and, therefore, we explicitly avoid
systematic correlations associated with the fragment sizes.
A large concentration of points close to the vertices indicates a fragmentation mode in which the size distribution has
a big fragment whereas the others are appreciably smaller.
Three big fragments of approximately the same size lead to points grouped around the center of the triangle.
On the other hand, when two fragments have approximately the same size, but are much larger than the third one, one finds
points gathered close to the middle point of the triangle's sides.

\begin{figure}[tb]
\includegraphics[width=5.5cm,angle=0]{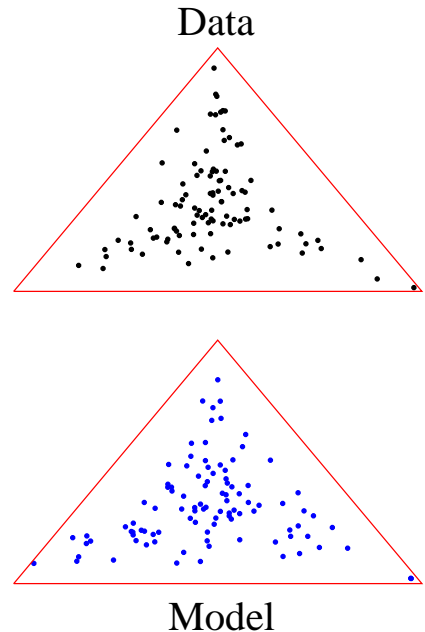}
\caption{\protect \label{fig:DalitzA} (Color online) Dalitz plot of the area distribution associated with the fragments produced
in the fragmentation of the alumina plates.
For details, see the text.}
\end{figure}

The experimental Dalitz plot displayed in Fig.\ \ref{fig:DalitzA}, obtained in the fragmentation of the alumina plates,
show that there is an important contribution from events in which one fragment is much larger than the others.
The results shown in this figure correspond to the cylinder pressure equal to 14 bar, but the conclusion remains valid
for the lower pressures employed in this work.
The corresponding model results are also displayed in the lower part of this figure and the qualitative agreement
with the experimental features is, once more, very good.

\begin{figure}[tb]
\includegraphics[width=5.5cm,angle=0]{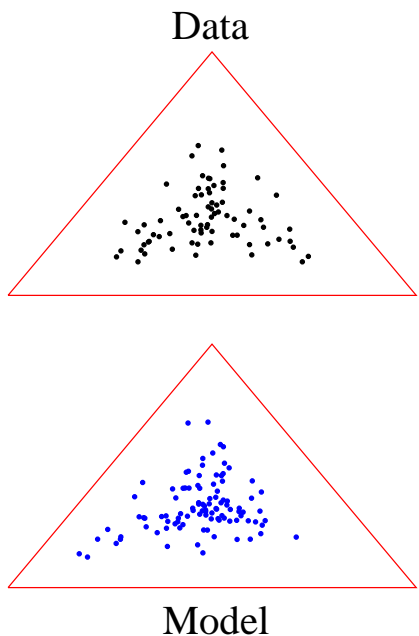}
\caption{\protect \label{fig:DalitzG} (Color online) Same as fig.\ \ref{fig:DalitzA} for the glass plates.
For details, see the text.}
\end{figure}

A different behavior is observed in the case of the glass plates, whose Dalitz plot tends to give points which are grouped
near the center of the triangle, for all the pressures we considered.
As discussed above, this means that the three largest fragments have approximately the same area.
Figure \ref{fig:DalitzG} shows the corresponding experimental and theoretical plots.
Due to the low experimental statistics for this kind of plot, the data associated with different pressures are grouped in this
figure, which does not affect our analysis as they are very similar for all pressures.
The model correctly predicts the tendency to suppress contributions associated with a dominant fragment.
Therefore, the agreement with the data is also fairly good in this case.

Further insight into the properties of the fragment distribution can be obtained by examining the shape parameter $Q$,
that we define as:

\begin{figure}[tb]
\includegraphics[width=7.0cm,angle=0]{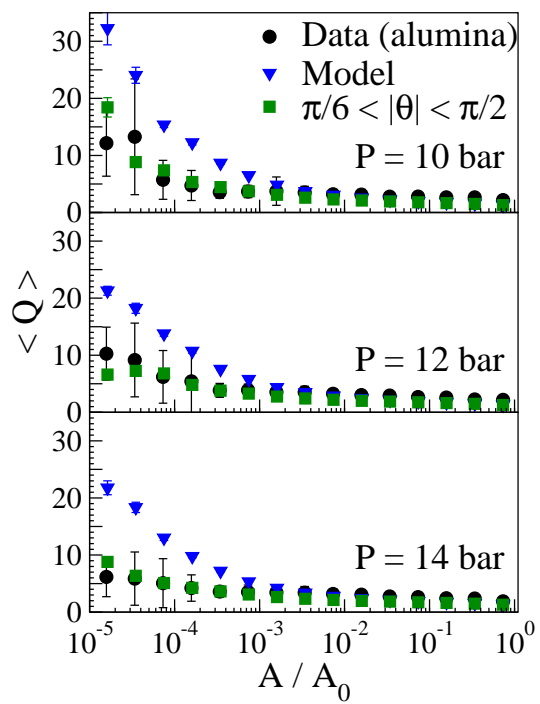}
\caption{\label{fig:q_al} (Color online) Average $Q$ value versus the fragments' area for different values of the
pressure for the fragmentation of the alumina plates.
For details, see the text.}
\end{figure}

\begin{equation}
Q=\frac{P^2}{4\pi A}\;,
\label{eq:q}
\end{equation}

\noindent
where $P$ denotes the perimeter of the fragment and $A$ corresponds to its area.
Large $Q$ values indicate elongated shapes whereas approximately circular or squared shapes give $Q$ close to unity.
In this way, this quantity allows one to obtain information on the shape of the fragments.
It is worth mentioning that we analyzed the fractal dimension of the borders and found that it is essentially 
one dimensional.
Figure \ref{fig:q_al} displays the average value of $Q$ for different impact velocities, obtained in
the fragmentation of the alumina plates.
It shows that the small fragments tend to be fairly elongated but this tendency is quickly weakened as the
violence of the impact increases.
On the other hand, the not too small fragments, i.e. $A/A_0 \gtrsim 10^{-4}$, are much less elongated
than the small ones, for all the pressures studied in this work.
The understanding of this property requires the development of a model that describes the fracture process accurately.
In this context, it should also be interesting to investigate whether there exists a preferential direction
to the elongation, but this is beyond the scope of this study as the present experimental setup does not provide
this kind of information.

\begin{figure}[bt]
\includegraphics[width=7.0cm,angle=0]{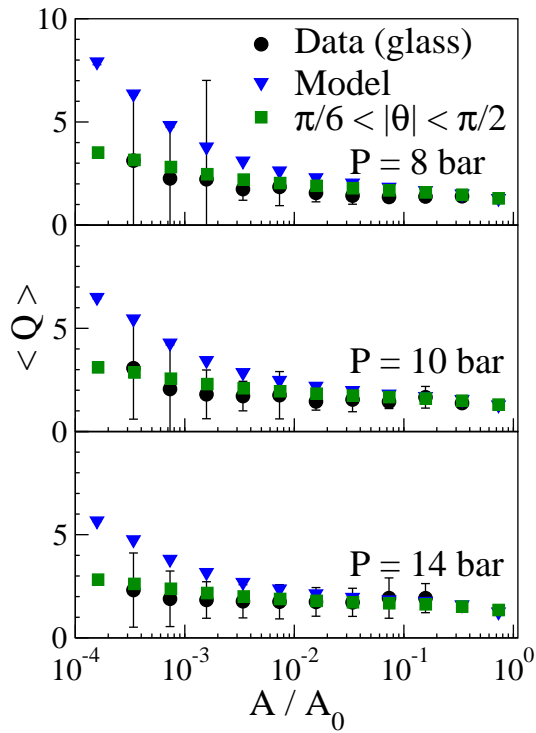}
\caption{\label{fig:q_g} (Color online) Same as Fig.\ \ref{fig:q_al} for the glass plates.}
\end{figure}

The average value of $Q$ obtained with our model is also shown in Fig.\ \ref{fig:q_al} and is depicted by the upside-down triangles.
The agreement with the experimental values is fairly good for $A/A_0\gtrsim 10^{-3}$, but the model predicts too elongated
fragments for smaller areas.
We come back to this point below.

The fragments produced in the breakup of the glass plates have different properties since the $\langle Q\rangle$
versus $A$ curve is fairly flat, except at the lowest pressures, as shown in Fig.\ \ref{fig:q_g}.
It should be noticed that the vertical scales in Figs. \ref{fig:q_al} and \ref{fig:q_g} are not the same.
More specifically, the elongation of the fragments originated from the glass plates is much smaller than that
observed in the case of the alumina objects.
Despite the experimental uncertainties, it is clear that the model, once more, systematically predicts too elongated fragments as already
noted in the case of the alumina plates.
These results suggest that the observable $Q$ might be a useful tool to selecting models which
give an appropriate picture to the fragmentation process.

The tendency to form too elongated fragments in our model can be appreciably reduced
by not allowing the cracks initiated at the flaw points to propagate in directions close to the original crack.
We have checked that the quality of the agreement with all the observables presented above is maintained if the bifurcation
angle $\theta$ is, for instance, restricted to $\pi/6 < \mid\theta\mid < \pi/2$.
All the other parameters keep their values, except for $P_c$ which is slightly changed (by less than 4\%) in some cases.
To illustrate this fact quantitatively, the $Q$ values obtained in this case are depicted by the squares in Figs.\ \ref{fig:q_al}
and \ref{fig:q_g}.
It is clear that the agreement with the data improved considerably.
Since there are experimental and theoretical studies \cite{dynCrack2003,dynCrackYoffe,dynCrack2,dynCrack3}
that give support to this angular restriction, our results strongly suggest that the new cracks start at relatively
large bifurcation angles.
In spite of these encouraging results, we preferred not to tune this model parameter and we keep this conclusion
on a qualitative level.
We postpone detailed discussions on this subject to future work when further observables will be analyzed.
In this work we intend to stress that models can reproduce global quantities, such as the mass distribution or the average fragment
multiplicity, whereas they can fail in describing more detailed information on the fragments' properties and that
important physical aspects can be revealed by restricting the model parameters.

As a final remark, it would be fair to speculate whether the experimental $\langle Q\rangle$ values, for small areas,
are not biased by the fact that we discard fragments whose areas are smaller than $A_{\rm cut}$.
It should be noticed that this lower limit is a safe cut which is much larger than the image resolution.
The specific values of $A_{\rm cut}$ have been selected for the alumina and glass plates based on the criterion
that the dimensions of the fragments are larger than the thickness of the plates due to the ambiguity associated
with the identification of the corresponding dimensions.
Although smaller fragments could be identified, they have been excluded
from the actual calculation of $\langle Q\rangle$ and other observables.
Qualitative analyses of such small fragments indicate that our results should not be impacted by the
consideration of these fragments.

\section{\label{sec:conclusions}Concluding Remarks\protect}
In summary, we have presented the results of an experiment in which the fragmentation of two different brittle materials due to
lateral impact was studied.
In agreement with former experimental works, we found that the size distribution is given by a power law over a wide range
of sizes.
The power law exponents are very similar for both materials, i.e. glass and alumina, and respectively correspond to
$\beta\approx 1.2$ and and $\beta\approx 1.1$, which are close to the values obtained in Refs. \cite{ODDERSH,fragJapanese,drops,PLATESprl}.

Going beyond those studies, the present experiment also focused on the geometric properties of the fragments.
The description of these properties provide strict tests to theoretical models. 
More specifically, we found that the size distribution and average fragment multiplicities are very well described by the simple
fragmentation model presented here, whereas it failed to reproduce the average elongation of the fragments' shapes.
This quantity is reproduced by the model only if the bifurcation angle of the fractures is restricted to relatively large angles ($\pi/6<\mid\theta\mid<\pi/2$) with respect to the propagation of the initial crack.
This is in qualitative agreement with the experimental results reported in Ref.\ \cite{dynCrack2003}
and the theoretical studies discussed in Refs.\ \cite{dynCrackYoffe,dynCrack2,dynCrack3}.
We believe that careful comparisons of these experimental observables \footnote{The experimental data presented in this work is available upon request.} with the predictions of different models
might be very useful in helping to establish a clear scenario for the fragmentation process.

\begin{acknowledgments}
We would like to acknowledge CNPq, CAPES, FAPERJ BBP grants, CNPq-PROSUL, and the joint PRONEX initiative of CNPq and FAPERJ under
Contract No.\ 26.171.528.2006, for partial financial support.
\end{acknowledgments}

\bibliography{manuscript}

\end{document}